\documentclass[aps,prl,showpacs,twocolumn]{revtex4}
\usepackage{epsfig,amsmath,color,amssymb}
\usepackage[10pt]{moresize}

\begin{document}

\title{Transverse Mode Revival of a Light-Compensated Quantum Memory}
\author{Fan Yang$^1$}
\author{Torsten Mandel$^1$}
\author{Christian Lutz$^1$}
\author{Zhen-Sheng Yuan$^{1,2}$}
\author{Jian-Wei Pan$^{1,2}$}
\affiliation{$^1$Physikalisches Institut,
Ruprecht-Karls-Universit\"{a}t Heidelberg, Philosophenweg 12, 69120
Heidelberg, Germany}
\affiliation{$^2$Hefei National Laboratory for
Physical Sciences at Microscale and Department of Modern Physics,
University of Science and Technology of China, Hefei, Anhui 230026,
China}
\date{\today}

\begin{abstract}
A long-lived quantum memory was developed based on light-compensated
cold $^{87}$Rb atoms in a dipole trap. The lifetime of the quantum
memory was improved by 40 folds, from 0.67 ms to 28 ms with the help
of a compensation laser beam. Oscillations of the memory efficiency
due to the transverse mode breathing of the singly-excited spin wave
have been clearly observed and clarified with a Monte-Carlo
simulation procedure. With detailed analysis of the decoherence
processes of the spin wave in cold atomic ensembles, this experiment
provides a benchmark for the further development of high-quality
quantum memories.
\end{abstract}


 \maketitle

Quantum information processing (QIP) has promising advantages over
classical information processing from the aspects of security,
channel capacity and computing efficiency \cite{NielsenQCQI2000,
KnillNature2001, GisinRMP2002, PanArxiv2008}. Therefore, one may
expect that a quantum computer could solve a physical problem, say
exploring the property of some new functional material, within a few
seconds for which it takes several years with a classical computer.
However, such a quantum computer has not yet been developed since
the current QIP systems are not scalable which means the required
resources increase exponentially along the complexity of the system
\cite{BriegelPRL1998}. This situation is going to be changed by
integrating the quantum memory, a storage device for quantum states,
into the QIP systems. For example, the concept of quantum repeater
\cite{BriegelPRL1998} has attracted much attention especially when
the atomic-ensemble-based quantum memory was proposed to be
integrated in the linear optical quantum network
\cite{DuanNature2001, ZhaoPRL2007, JiangPRA2007, CollinsPRL2007},
where the required resources increase polynomially along the
communication distance. This means the QIP systems become scalable
when quantum memories are integrated in \cite{KimbleNature2008,
YuanPR2010}.

Along the direction of developing high-quality quantum memories,
great progress has been achieved in the recent years. Various
mediums have been adopted for quantum memories, e.g. atomic
ensembles \cite{MatsukevichSCI2004,
EisamanNature2005,ChenNPhys2008}, single atoms
\cite{MoehringNature2007, RosenfeldPRL2007}, ion-doped crystals
\cite{AfzeliusNature2008, LongdellPRL2005}. Physical processes of
electromagnetically induced transparency (EIT)
\cite{EisamanNature2005, LongdellPRL2005}, Raman scattering
\cite{MatsukevichSCI2004, ChenNPhys2008}, and photon echo
\cite{AfzeliusNature2008} are involved in the manipulation of
quantum memories. Along the way for long-distance quantum
communication network, till now the most advanced achievement is the
realization of quantum repeater nodes \cite{ChouScience2007,
YuanNature2008} based on cold-atomic-ensemble quantum memories. The
memory performance, measured by its lifetime and retrieve
efficiency, determines the complexity of a QIP system. Therefore
continuously pushing the upper-limit of the lifetime and retrieve
efficiency is indispensable for the future scalable QIP systems.

For the single-photon quantum memory with cold atomic ensembles, a
lifetime of milliseconds was achieved in our former experiment by
suppressing the dephasing with a long wavelength of the spin wave
\cite{ZhaoboNPHYS2009} and in another independent experiment by
limiting the motion of the atoms with a one-dimensional optical
lattice \cite{ZhaoranNPHYS2009}. The major sources of decoherence of
these quantum memories are the atom loss due to thermal diffusion
and gravity when the cold atoms are freely flying in the vacuum
chamber \cite{ZhaoboNPHYS2009}, or the differential light shift when
the cold atoms are confined with an optical trap
\cite{ZhaoranNPHYS2009}. This can be understood when we look at the
formula of the generated spin wave,
\begin{eqnarray}\label{eqn:collective}
|\psi\rangle_c= \frac{1}{\sqrt{N_\textrm{\tiny
at}}}\sum_j^{N_\textrm{\tiny at}} e^{i\omega^{hf}_j\cdot t}
e^{i\Delta\boldsymbol{k} \cdot \boldsymbol{r}_j}|g \cdots s_j \cdots
g \rangle.
\end{eqnarray}
Here, $N_\textrm{\tiny at}$, $\omega^{hf}_j$,
$\Delta\boldsymbol{k}$, and $\boldsymbol{r}_j$ are the number of the
atoms, the hyperfine split of the $j$th atom, the momentum transfer
from light to atoms, and the position of the $j$th atom
respectively. The two phases $\phi_1=(\omega^{hf}_j\cdot t)$ and
$\phi_2=(\Delta\boldsymbol{k} \cdot \boldsymbol{r}_j)$ are
correlated with the differential light shift and the motion of the
atoms respectively and have to be maintained for keeping the phase
pattern.

\begin{figure}[tb]
   \centering
   \includegraphics[width=0.45\textwidth]{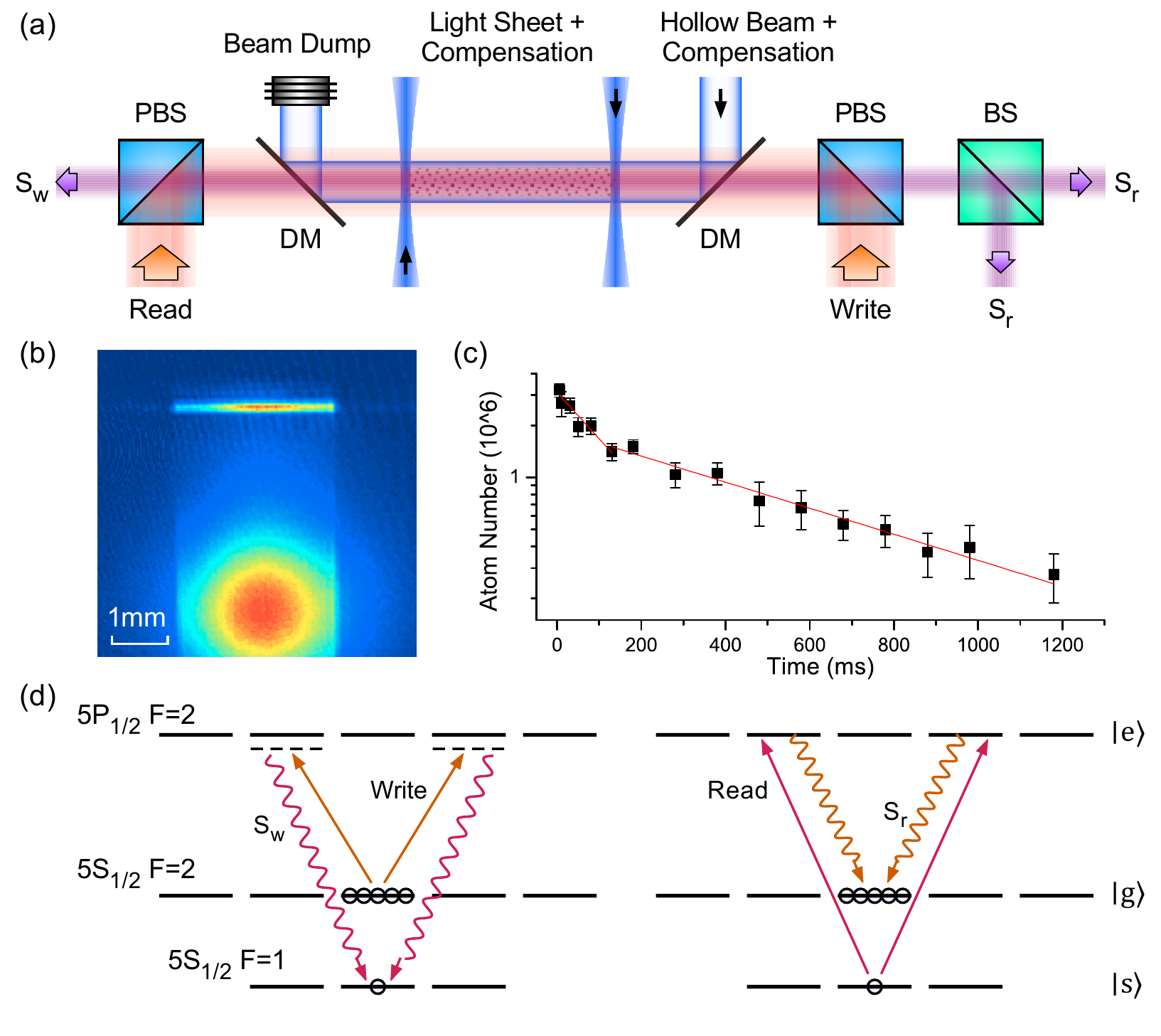}
\caption{A schematic of the setup. (a) The memory light modes are
arranged in a collinear configuration, along the axial direction of
the hollow beam. The light modes of the write, read, write-out
photon $S_w$, and read-out photon $S_r$ are all focused single-mode
Gaussian beams. At the focus, the diameter for the write and read is
550 $\mu$m and that for the signal beams is 130 $\mu$m. (b) An
absorption image of the atoms in the dipole trap (with false color).
The dipole trap holds $\sim 3\times10^6$ atoms at $30~\mathrm{ms}$
after sub-Doppler cooling, in a shape with length $3~\mathrm{mm}$
and diameter $190~\mathrm{\mu m}$. (c) The loss of atoms in the
optical trap. Within the first $100~\mathrm{ms}$, the hot atoms
quickly escape from the trap, and an exponential fitting gives the
decay constant of $160~\mathrm{ms}$. After that, the atom number
slowly decays with a constant of $580~\mathrm{ms}$. (d) Relevant
atomic levels. A pair of clock states, $|F=2,m_F=0\rangle$ and
$|F=1,m_F=0\rangle$,  are chosen as the memory states to suppress
dephasing from inhomogeneous magnetic field. BS, 50:50 beam
splitter; PBS, polarizing beam splitter; DM, dichroic mirror.}
\label{fig:setup}
\end{figure}

In order to eliminate these decoherence effects in our former
experiment \cite{ZhaoboNPHYS2009}, in the present experiment we have
developed a blue-detuned optical trap to confine the cold atoms and
a weak light field to compensate the differential light shift in
light of the idea in Ref. \cite{KaplanJOB2005}. As shown in the
following text an improvement of the lifetime of the spin wave by 40
folds, from 0.67 ms to 28 ms, has been obtained with the help of the
compensation beam. Moreover, the transverse mode breathing of the
singly-excited spin wave is observed for the first time and clearly
interpreted by a Monte-Carlo simulation procedure.

\begin{figure}[tb]
   \centering
   \includegraphics[width=0.33\textwidth]{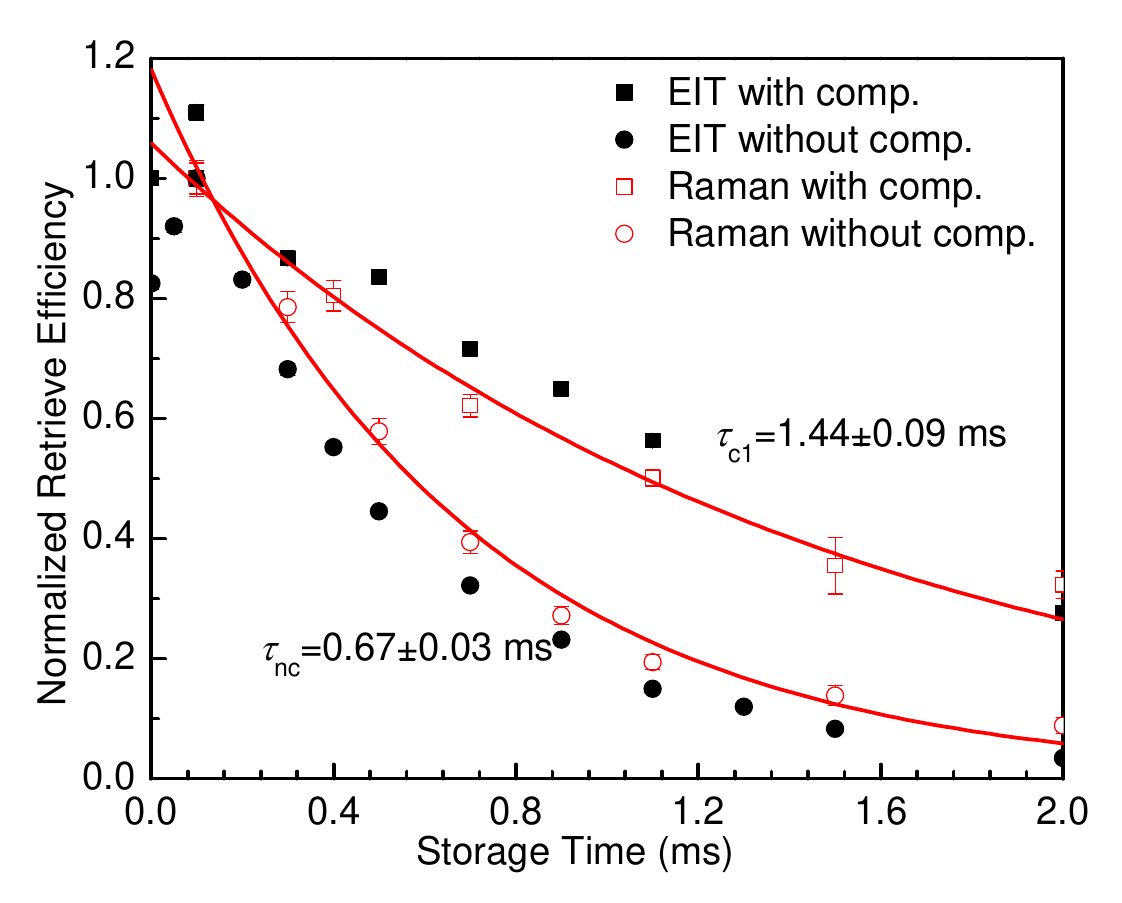}~
   \includegraphics[width=0.13\textwidth]{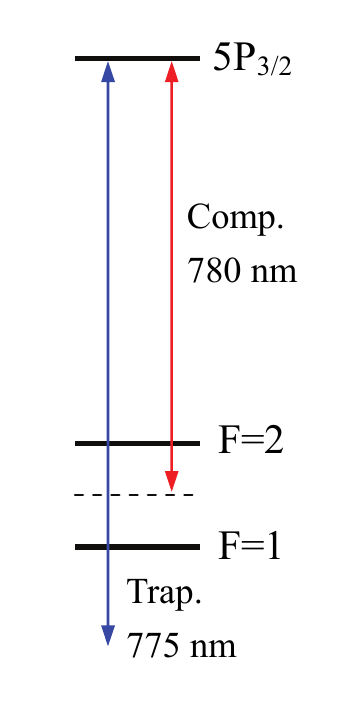}\\
~~~~~~~~~~~~~~~~~~~(a)~~~~~~~~~~~~~~~~~~~~~~~~~~~~~~~~~(b)
\caption{(a) Retrieve efficiency at short time scale. (b) The
wavelength of the trapping beam is 775 nm and that for that of the
compensation beam is 780 nm. Without compensation, dominated
dephasing mechanism is differential light shift from the trapping
laser, and an exponential fitting gives a lifetime of
$0.67\pm0.03~\mathrm{ms}$. When the compensation laser is turned on,
the decay constant becomes $1.44\pm0.09~\mathrm{ms}$.}
\label{fig:short}
\end{figure}

Shown in Fig.\ref{fig:setup}(a) the blue-detuned optical trap
consists of a hollow beam and two light sheets both delivered from a
Ti:Sapphire laser working at wavelength 775 nm and a power of 1.9 W.
The hollow beam is generated by the so-called convex-axicon
combination \cite{inPrep}, which has a steep ascent of laser
intensity along the radial direction. The focal length of the convex
lens is 150 mm, and the edge angle of the axicon lens is
$0.073^{\circ}$. Taking a collimated Gaussian mode with a diameter
of 16 mm as the input, the laser beam is transformed to a
ring-shaped profile of 190-$\mathrm{\mu m}$ inner diameter at the
focus. The 3-mm long cylindrical section around the focus is taken
as the trap region. Another two light sheets with elliptical profile
close the two ends of the optical trap. Unlike the red-detuned
single-beam optical trap \cite{ChuuPRL2008} where cold atoms are
collected at the intensity maximum of the trapping beam, in this
blue-detuned ``box''-like trap cold atoms stay in the dark region.
Such a trap has the advantages of lower scattering rate and smaller
differential light shift over the red-detuned trap when they have
the same trap depth and comparable detuning. Moreover, it is more
feasible to adjust the inner diameter of the hollow beam to match
the optical modes of the signal beams $S_w$ and $S_r$.

During the experiment, the dipole trap passing through the center of
a dark magneto-optical trap (dark MOT) \cite{inPrep} is kept on all
the time. Around $3 \times 10^6$ cold $^{87}$Rb atoms at a
temperature of 15 $\mu$K are transferred into the dipole trap after
a 150-ms dark MOT phase  and a 6-ms sub-Doppler cooling phase. Then,
a weak magnetic field of 340 mG  is turned on along the axial
direction of the hollow beam, which defines the quantization axis.
The atoms outside the dipole trap fall down and expand from the
trapped atoms as shown in Fig.\ref{fig:setup}(b) by waiting for
another $30~\mathrm{ms}$. Afterwards within 100 $\mu$s the trapped
atoms are optically pumped to the initial state $|g\rangle$ with a
measured optical depth of $9.6\pm1.0$ along the quantization axis.
The states of $|g\rangle$ and $|s\rangle$, a pair of ``clock''
states, are taken as the memory states. From now, the memory cycles
repeat for 10 to 300 ms (depending on the storage time) until a new
trap loading cycle starts.

In the present experiment, both the off-resonance Raman scattering
and the EIT processes are utilized to test the quality of the
quantum memory. When the Raman scattering is used to generate spin
excitations via $|g\rangle\rightarrow|e\rangle\rightarrow|s\rangle$,
the excitation rate is controlled to about 2$\times10^{-3}$. After a
photon is detected in the signal mode $S_w$, it triggers the read
pulse at a storage time $\Delta T$ to convert the collective
excitation into a single photon in the mode of $S_r$. The retrieve
efficiency is estimated by the ratio between the count in the $S_r$
channel and that in the $S_w$ channel. When we study the EIT
process, the control beam with a power of 200 $\mu$W is sent in from
the read channel and the probe beam is coupled in from the $S_w$
channel and detected at the $S_r$ channel. The intensity of the
probe pulse is set at single photon level. The memory efficiency,
optimized according to an iterating procedure
\cite{NovikovaPRL2007}, is defined by the fraction being read out
relative to the intensity of the probe pulse itself.

Since the power of the trapping laser is about 1.9 W, its wavelength
is set to 775 nm to maintain a trap depth of  45 $\mu$K. The
estimated coherence time is 0.85 ms due to the differential light
shift. A compensation laser with power of 3.5 $\mu$W and frequency
tuned between the two DII resonances (shown in
Fig.\ref{fig:short}(b)) is mixed into the trapping laser with a beam
sampler before they are coupled into a single-mode
polarization-maintaining fiber. After propagating in the fiber,
these two lasers share an identical profile. Taking the chromatic
abberation of the following optics into account, the estimated
lifetime is $\sim 300~\mathrm{ms}$. The two lasers are frequency
stabilized and the relative intensity drift between them is actively
controlled to smaller than $1\%$.

The retrieve efficiencies measured with the Raman scattering and the
EIT process are shown in Fig.\ref{fig:short}(a). The detected
retrieve efficiency at a storage time below 100 $\mu$s is about 7\%
when the Raman scattering is utilized. By removing the
inefficiencies of mode-matching, transmittance loss, and detection
efficiency of the single photon counter module, the intrinsic
retrieve efficiency is about $\eta_{\mbox{\tiny RS}}$=30\%. The
memory efficiency with the EIT process is obtained as
$\eta_{\mbox{\tiny EIT}}$=22\% where the inefficiencies have been
automatically removed during the measurement. These two efficiencies
($\eta_{\mbox{\tiny RS}}$ and $\eta_{\mbox{\tiny EIT}}$) are well
comparable. In order to compare the efficiency curves, the data
points are normalized to the value at short storage times. It can be
seen that decay curves obtained from the two physical processes are
consistent to each other. Meanwhile, in order to show the single
photon nature of the retrieved photon from the spin wave generated
with off-resonance Raman scattering, we measured the
anti-correlation parameter \cite{ChenShuaiPRL2006} as $0\pm0.1$ at
1.2 ms.

\begin{figure}[tb]
\centering
\begin{tabular}{ll}
\raisebox{30ex}{(a)}& \includegraphics[width=0.40\textwidth]{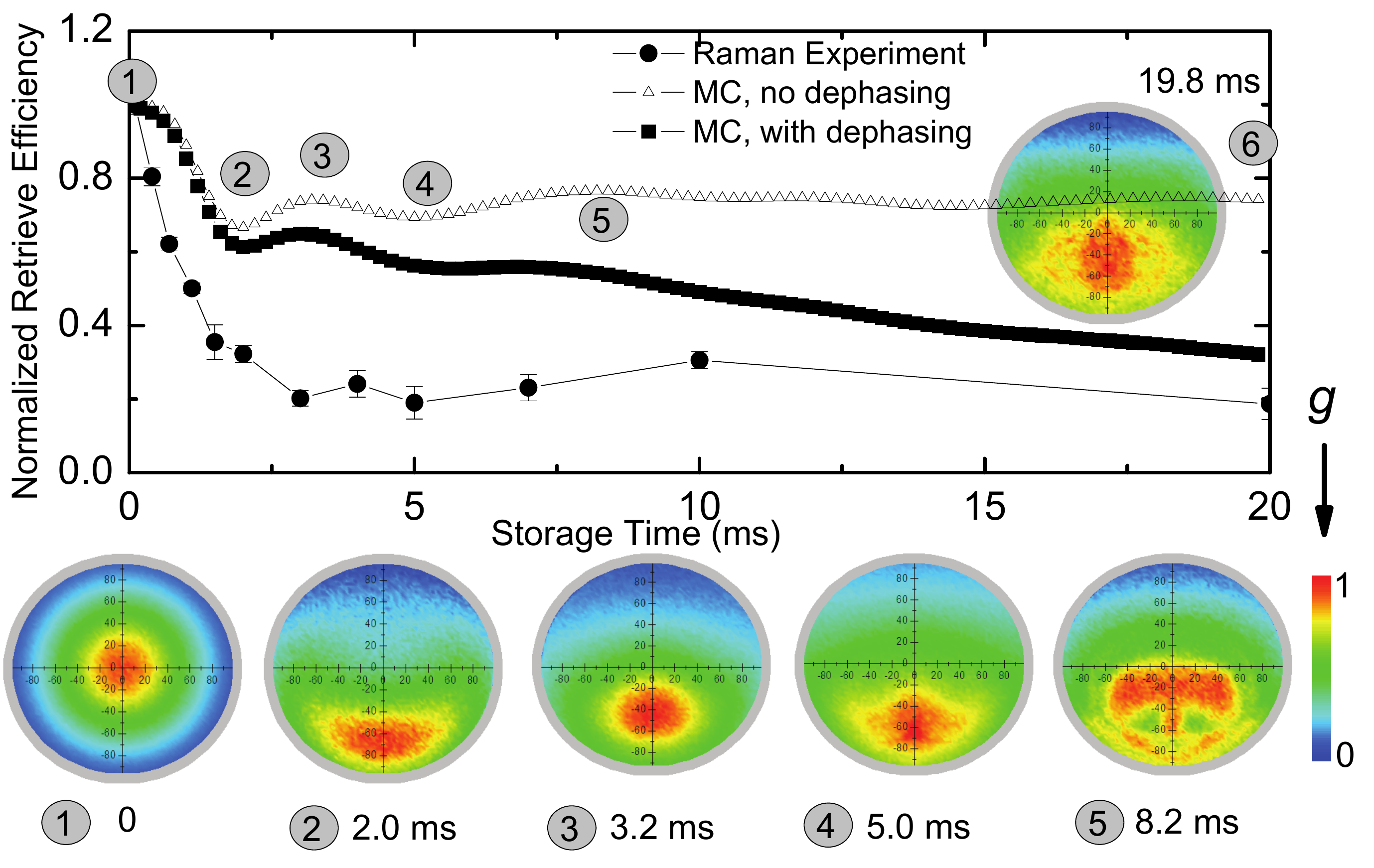}\\
\raisebox{30ex}{(b)}&
\includegraphics[width=0.40\textwidth]{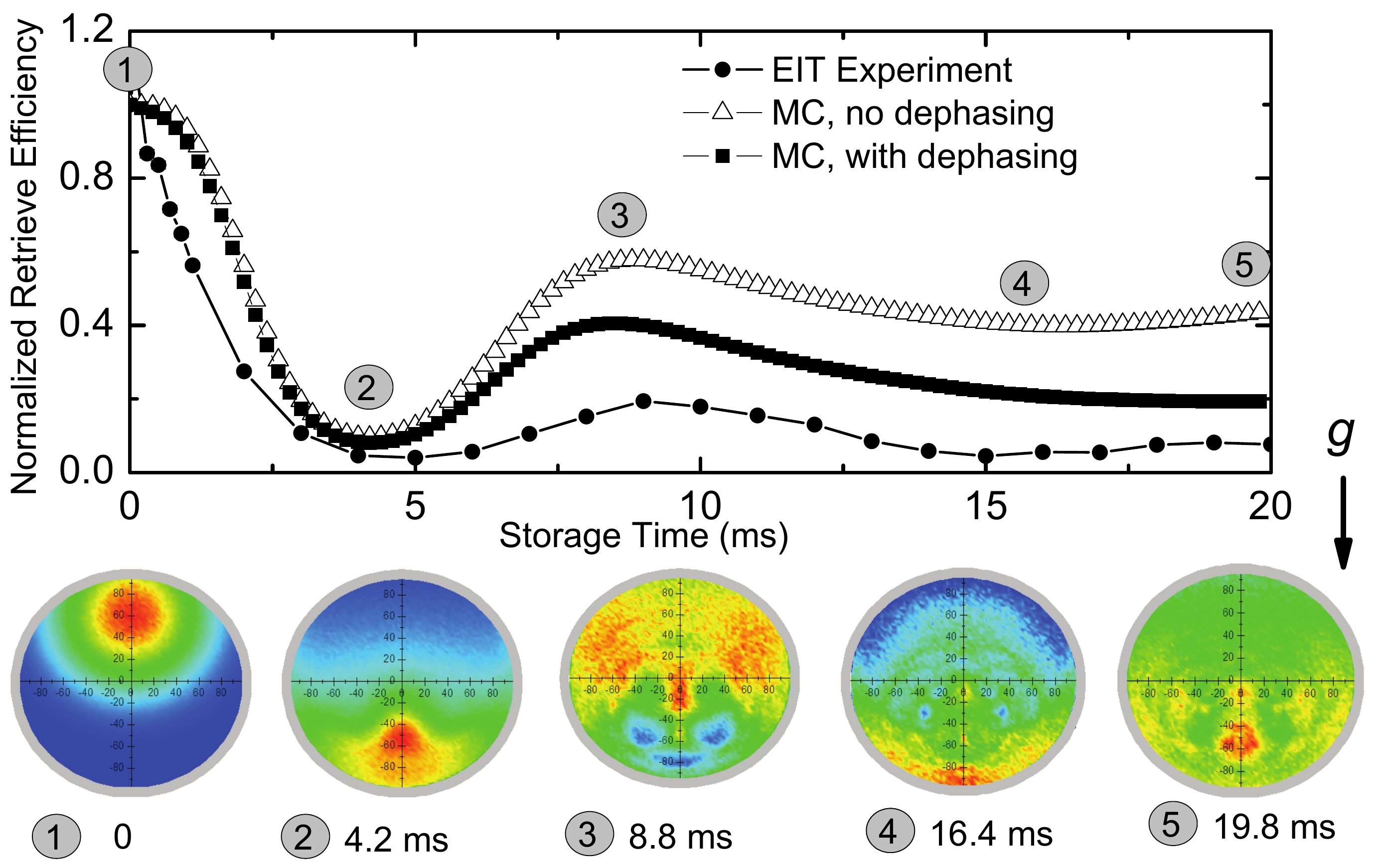}
\end{tabular}
\caption{Oscillation of retrieve efficiency due to dynamical mode
breathing. (a) Retrieve efficiency curve when the signal mode is at
the center of the trap and the spin wave is generated by
off-resonance Raman scattering. (b) Retrieve efficiency curve when
the signal mode is 60 $\mu$m above the center of the trap and the
spin wave is generated by the EIT process. The transverse spatial
mode of the singly-excited spin wave at different storage times is
simulated with a Monte-Carlo procedure and shown below the retrieve
efficiency curves. The triangle points are the overlapping integral
calculated directly from the initial spatial mode and the evolved
spatial mode. The squared dots are obtained by considering atom loss
and the dephasing time obtained in Fig.\ref{fig:long}. The color
depth shows the density of the atoms and is re-scaled for each
figure to make the distribution more visible.} \label{fig:rev}
\end{figure}

The fitted lifetime without the compensation beam is 0.67$\pm$0.03
ms which is well in agreement with the estimated value of 0.85 ms.
However, the decay constant with the compensation beam is
1.44$\pm$0.09 ms which is far from the estimated value of 300 ms.

For the collinear phase-matching configuration as in
Fig.\ref{fig:setup}(a), the generated spin-wave has a wavelength
$\sim 4.4~\mathrm{cm}$, which is much longer than the trap length 3
mm. Therefore the dephasing from atomic thermal motion along the
axial direction is suppressed and the induced decoherence is
negligible. When the measurement is taken at longer storage times,
obvious oscillations emerge in the curves of the memory efficiencies
as in Fig.\ref{fig:rev}.

We first look at Fig.\ref{fig:rev}(a) which was measured with the
Raman scattering. Two dips at 3 ms and 5 ms and two maxima at 4 ms
and 10 ms are observed. The density of the atoms is about
$5\times10^{10}$ cm$^{-3}$, which should not induce the spin self
rephasing at several milliseconds due to the identical spin rotation
effect observed in a former experiment \cite{DeutschPRL2010}. In
order to explore the mechanism for the phase revival, we construct a
model for numerically simulating motions of the atoms in the trap
with a Monte-Carlo procedure.

When a single photon is registered in the $S_w$ channel which has a
Gaussian profile defined by the single-mode fiber and its
collimator, the spin wave possesses a Gaussian profile accordingly
with a initial transverse distribution of $U(x,y,t=0)$. After a
period of thermal expansion and free falling, the transverse spatial
mode of the spin wave evolves into $U(x,y,t=\Delta T)$. Then we
define the retrieve efficiency as the overlap of the two modes, i.e.
$R(\Delta T)=|\int\int\mathrm{d}x\mathrm{d}y \sqrt{U(x,y,0)}
\sqrt{U(x,y,\Delta T)}|^2$. As the two spatial modes $U(x,y,0)$ and
$U(x,y,\Delta T)$ are identical at $\Delta T=0$, their overlap is
unit. At $\Delta T=2.0$ ms, the atoms get more diffused and fall
down to the lower part of the trap. Then a dip appears in the
efficiency curve due to the decreasing overlap. At $\Delta T=3.2$
ms, the atoms are bounced higher and more concentrated. Therefore
the overlap becomes larger and a maximum arises. At $\Delta T=5.0$
ms, the atoms get diffused again and the overlap becomes smaller
therefore the second dip appears. Until $\Delta T=8.2$ ms, even
though the atoms expand more which should reduce the overlap, the
position of the atoms is closer to the initial position which should
increase the overlap. As a result of the competition between these
two effects, a flat maximum arises. After a sufficiently long time
until 19.8 ms, the distribution of the atoms becomes more chaotic
and the oscillations disappears. The features in the experimental
curve are reproduced in the Monte-Carlo simulation according to the
transverse mode breathing although the positions are slightly
shifted. This discrepancy may come from the deviation of the trap
diameter from 190 $\mu$m, the inhomogeneity of the trap and the
imperfection of the simulation model.

\begin{figure}[tb]
   \centering
   \includegraphics[width=0.4\textwidth]{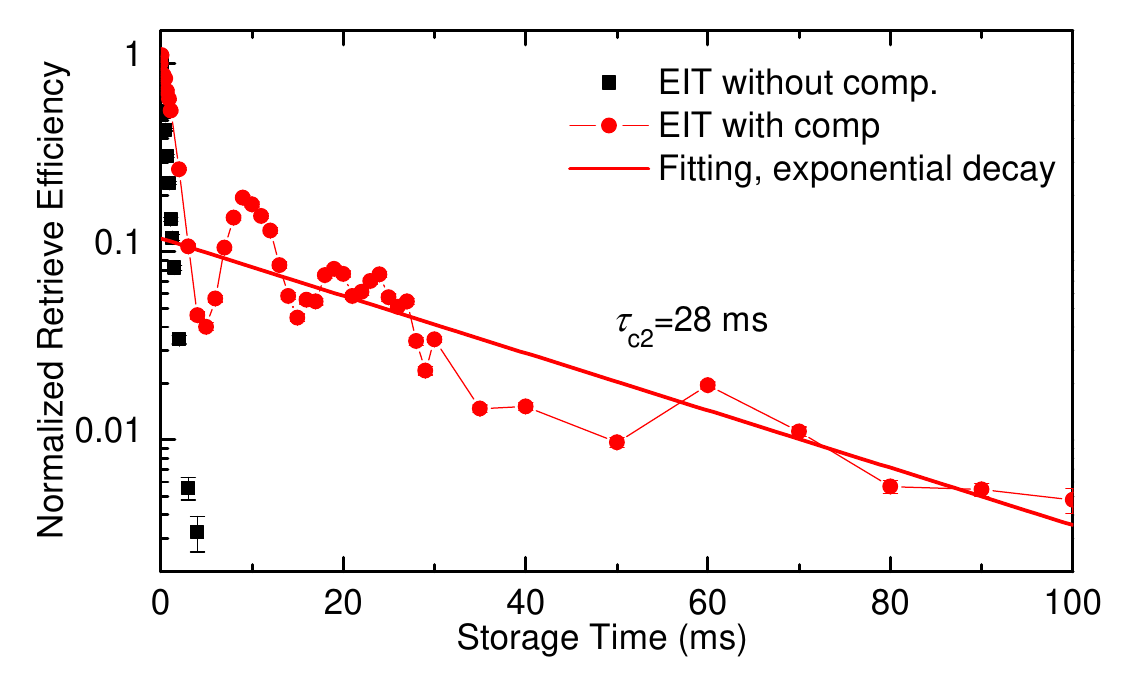}
\caption{Memory efficiency at long time scale. By a least-square
fitting with an exponential function to the measured data, a decay
constant of 28$\pm$2.5 ms is obtained. }
   \label{fig:long}
\end{figure}

In order to make this dynamical mode breathing effect more visible,
we shifted the signal beam 60 $\mu$m above the trap center. As shown
in Fig.\ref{fig:rev}(b), the oscillations of memory efficiency
become much more evident. The features of the revivals in the
experimental result are all reproduced in the Monte-Carlo
simulations. As explained above, the discrepancies between the
experimental curve and the simulated results should arise from the
similar reasons.

At even longer time scale, the memory efficiency curve is measured
up to 100 ms. By a least-square fitting to the experimental data
with an exponential function, a lifetime of 28$\pm$2.5 ms is
obtained, which indicates an improvement by 40 folds. However, it is
still much shorter than the estimated value of 300 ms. This is due
to the spatial mode mismatching between the trap beam and the
compensation beam arising from the imperfection of the optics.
Moreover, the relative intensity stability of the compensation beam
of 1\% limits the compensation effect as well.

In summary, the present work showed a 40-fold improvement of the
coherence time of the quantum memory based on optically trapped
atoms with a compensation laser beam. Especially, the transverse
mode breathing effect, a collective behavior of the non-interactive
atoms, of a singly-excited spin wave was observed and elucidated
with a Monte-Carlo simulation procedure. Since this effect is
induced by the transverse mode diffusion and revival instead of the
evolution of the phase pattern of the spin wave, it is not likely to
be observed with the traditional Ramsey spectroscopy. The present
experiment suggests that a three dimensional optical lattice is
necessary for completely suppressing the spatial mode diffusion of
atoms. With detailed analysis of the decoherence mechanisms, this
work provides a benchmark for the further development of
high-quality quantum memories.

This work was supported by the European Commission through the ERC
Grant and the STREP project HIP, the CAS, the NNSFC and the National
Fundamental Research Program (Grant No. 2011CB921300) of China.

\textit{Note added}: During the course of our experiment, we learnt
a parallel effort for extending the lifetime of a quantum memory by
about one order with the technique of compensating the differential
light shift in a red-detuned one-dimensional optical lattice
\cite{RadnaevNatPhys2010}.


\begin{thebibliography}{10}

\bibitem{NielsenQCQI2000}
M.~A. Nielsen and I.~L. Chuang.
\newblock \emph{Quantum Computation and Quantum Information}.
\newblock {Cambridge University Press} (2000).

\bibitem{KnillNature2001}
E.~Knill, \emph{et~al.}
\newblock Nature \textbf{409}, 46 (2001).

\bibitem{GisinRMP2002}
N.~Gisin, \emph{et~al.}
\newblock Rev. Mod. Phys. \textbf{74}, 145 (2002).

\bibitem{PanArxiv2008}
J.-W. Pan, \emph{et~al.}
\newblock \textnormal{arXiv}:0805.2853  (2008).

\bibitem{BriegelPRL1998}
H.-J. Briegel, \emph{et~al.}
\newblock Phys. Rev. Lett. \textbf{81}, 5932 (1998).

\bibitem{DuanNature2001}
L.-M. Duan, \emph{et~al.}
\newblock Nature \textbf{414}, 413 (2001).

\bibitem{ZhaoPRL2007}
B.~Zhao, \emph{et~al.}
\newblock Phys. Rev. Lett. \textbf{98}, 240502 (2007).

\bibitem{JiangPRA2007}
L.~Jiang, \emph{et~al.}
\newblock Phys. Rev. A \textbf{76}, 012301 (2007).

\bibitem{CollinsPRL2007}
O.~A. Collins, \emph{et~al.}
\newblock Phys. Rev. Lett. \textbf{98}, 060502 (2007).

\bibitem{KimbleNature2008}
H.~J. Kimble.
\newblock Nature \textbf{453}, 1023 (2008).

\bibitem{YuanPR2010}
Z.-S. Yuan, \emph{et~al.}
\newblock Physics Reports \textbf{497}, 1  (2010).

\bibitem{MatsukevichSCI2004}
D.~N. Matsukevich and A.~Kuzmich.
\newblock Science \textbf{306}, 663 (2004).

\bibitem{EisamanNature2005}
M.~D. Eisaman, \emph{et~al.}
\newblock Nature \textbf{438}, 837 (2005).

\bibitem{ChenNPhys2008}
Y.-A. Chen, \emph{et~al.}
\newblock Nat. Phys. \textbf{4}, 103 (2008).

\bibitem{MoehringNature2007}
D.~L. Moehring, \emph{et~al.}
\newblock Nature \textbf{449}, 68 (2007).

\bibitem{RosenfeldPRL2007}
W.~Rosenfeld, \emph{et~al.}
\newblock Phys. Rev. Lett. \textbf{98}, 050504 (2007).

\bibitem{AfzeliusNature2008}
H.~de~Riedmatten, \emph{et~al.}
\newblock Nature \textbf{456}, 773 (2008).

\bibitem{LongdellPRL2005}
J.~J. Longdell, \emph{et~al.}
\newblock Phys. Rev. Lett. \textbf{95}, 063601 (2005).

\bibitem{ChouScience2007}
C.-W. Chou, \emph{et~al.}
\newblock Science \textbf{316}, 1316 (2007).

\bibitem{YuanNature2008}
Z.-S. Yuan, \emph{et~al.}
\newblock Nature \textbf{454}, 1098 (2008).

\bibitem{ZhaoboNPHYS2009}
B.~{Zhao}, \emph{et~al.}
\newblock Nat. Phys. \textbf{5}, 95 (2009).

\bibitem{ZhaoranNPHYS2009}
R.~{Zhao}, \emph{et~al.}
\newblock Nat. Phys. \textbf{5}, 100 (2009).

\bibitem{KaplanJOB2005}
A.~Kaplan, \emph{et~al.}
\newblock J. Opt. B-Quantum Semicl. Opt. \textbf{7}, R103 (2005).

\bibitem{inPrep}
F.~Yang, \emph{et~al.} (in preparation).

\bibitem{ChuuPRL2008}
C.-S. Chuu, \emph{et~al.}
\newblock Phys. Rev. Lett. \textbf{101}, 120501 (2008).

\bibitem{NovikovaPRL2007}
I.~Novikova, \emph{et~al.}
\newblock Phys. Rev. Lett. \textbf{98}, 243602 (2007).

\bibitem{ChenShuaiPRL2006}
S.~Chen, \emph{et~al.}
\newblock Phys. Rev. Lett. \textbf{97}, 173004 (2006).

\bibitem{DeutschPRL2010}
C.~Deutsch, \emph{et~al.}
\newblock Phys. Rev. Lett. \textbf{105}, 020401 (2010).

\bibitem{RadnaevNatPhys2010}
A.~G. Radnaev, \emph{et~al.}
\newblock Nat Phys \textbf{6}, 894 (2010).

\end{thebibliography}
\end{document}